\documentclass{Interspeech}

\interspeechcameraready

\title{Adaptive Knowledge Distillation for Device-Directed Speech Detection}

\author[affiliation={1, *}]{Hyung Gun}{Chi}
\author[affiliation={1, *}]{Florian}{Pesce}
\author[affiliation={1, *}]{Wonil}{Chang}
\author[affiliation={1, *}]{Oggi}{Rudovic}
\author[affiliation={1}]{\\ Arturo}{Argueta}
\author[affiliation={1}]{Stefan}{Braun}
\author[affiliation={1}]{Vineet}{Garg}
\author[affiliation={2, \dagger}]{Ahmed Hussen}{Abdelaziz}

\affiliation{}{Apple}{$^2$Meta}
\email{hchi23@apple.com}

\usepackage{comment}
\usepackage{multirow}
\usepackage{graphicx}
\usepackage{hyperref}
\usepackage{cite}
\usepackage{url}
\usepackage{xspace}
\newcommand{\ie}{i.e.\xspace}
\usepackage[capitalize]{cleveref}

\begin{document}

\maketitle

\renewcommand{\thefootnote}{}
\footnotetext{$^*$equal contribution.}
\footnotetext{$^\dagger$work done at while Apple.}
\begin{abstract}
Device-directed speech detection (DDSD) is a binary classification task that separates the user’s queries to a voice assistant (VA) from background speech or side conversations. This is important for achieving naturalistic user experience. To this end, we propose knowledge distillation (KD) to enhance DDSD accuracy while ensuring efficient deployment. Specifically, we introduce a novel adaptive KD method that transfers knowledge from general representations of an ASR large pre-trained acoustic encoder ({\it teacher}). We apply task-specific adapters, on top of the (frozen) teacher encoder, trained jointly with the student model on DDSD. We demonstrate that the proposed adaptive KD outperforms the student model without distillation in the keyword and keyword-free (follow-up) invocations, with an improvement of +26\% and +19\% in terms of Equal Error Rate, respectively. We also show that this approach generalizes across the transformer and conformer-based model architectures.
\end{abstract}

\section{Introduction}  \label{sec:intro}
Smart devices, such as mobile phones, wearables, and smart speakers, have become an integral part of our daily routines. This is facilitated by the use of VAs to enable a naturalistic user experience. Users can interact with these devices through various methods: voice-triggered commands using specific wake-words, touch-based inputs via physical or virtual buttons, and/on keyword-free follow-up modes. Device-directed speech detection (DDSD)  \cite{sainath2015convolutional, wu2018monophone, mallidi2018devicedirected, vineet_interspeech22} aims to distinguish between user queries directed at voice assistants, and background speech or side conversations. Thus, DDSD is crucial in preventing unintended activations, which can occur when speech resembles wake-words or accidental button presses, among others.

To improve the DDSD accuracy, we employ Knowledge Distillation (KD) \cite{hinton2015distilling, gou2021knowledge}, a technique that has demonstrated promising ability in compressing model sizes while mitigating performance trade-offs across various domains \cite{chang2022distilhubert, gandhi2023distil, muralidharan2024compact, asami2017domain, yang2020distilling, cui2017knowledge, chen2017learning, chen2022knowledge}. The motivation for applying KD to DDSD is two-fold: (i) the teacher is trained on a much larger corpus of audio data (compared to the audio data available for DDSD), and uses a larger model, tuned for the ASR task, producing more robust and general acoustic representations. (ii) Hardware requirements often constrain the model size and runtime memory; therefore devising a small yet accurate DDSD model is critical for on-device deployment. Multiple approaches in various domains such as computer vision \cite{chen2017learning, chen2022knowledge} and speech \cite{jang2023recycle, peng2023dphubert, gandhi2023distil, lee2022fithubert, chang2022distilhubert}, have been used for distillation from larger models to the smaller ones. However, little has been investigated when it comes to DDSD task~\cite{vineet_interspeech22}.

\begin{figure}[t]
    \centering
    \includegraphics[width=1.0\linewidth]{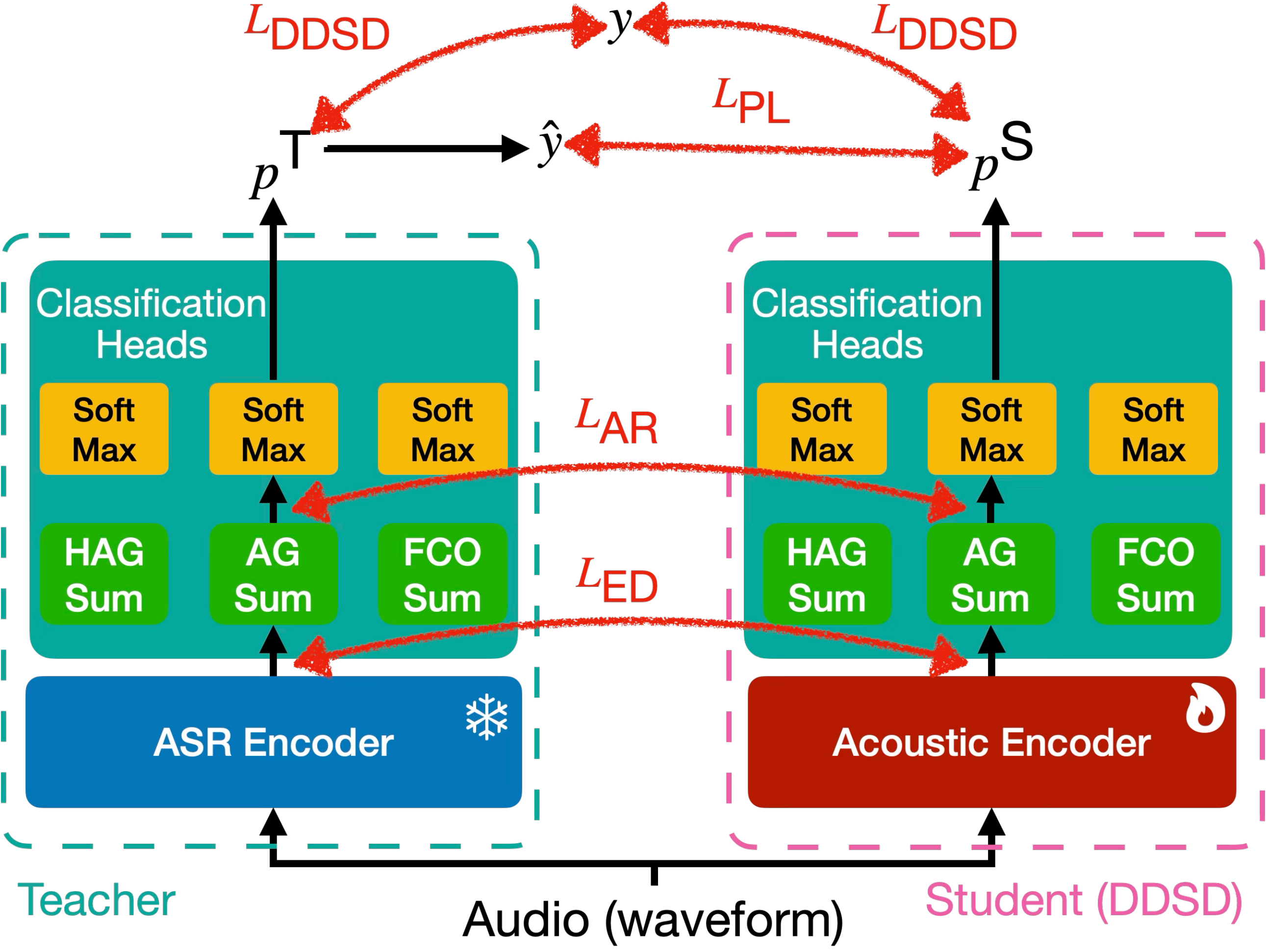}
    \caption{{\bf Adaptive knowledge distillation for on-device device-directed speech detection (DDSD)}. To bridge the domain gap, we add classification heads for the teacher model on top of a pre-trained ASR encoder. These teacher classification heads are jointly trained with the student on the DDSD task.}
    \vspace{-1em}
    \label{fig:main}
\end{figure}

The KD can be challenging, often requiring to fine-tune the teacher for the target task to narrow the domain gap (\ie data used to train the teacher and the task-specific data). However, fine-tuning large teacher models is resource-intensive. In this paper, we propose adaptive KD (aKD) for DDSD task, where the teacher/student adapters and the student encoder are trained simultaneously. This sets aKD apart from conventional KD, which first fine-tunes the teacher model for the target task, and then distills knowledge into the student from the frozen teacher. 

In aKD, we enable a dynamic alignment of the features between the student and teacher acoustic encoders, as well as their decision scores for each task - keyword detection ``Hey Agent" (HAG) and ``Agent" (AG), and follow-up conversations (FCO). This is depicted in Fig.~\ref{fig:main}, where the student and teacher are comprised of an acoustic encoder and classification heads ({\it adapters}). In this paper, the teacher encoder is a pre-trained (frozen) acoustic encoder for ASR~\cite{gulati2020conformer, yao2021wenet}; however, the aKD method is general enough so any foundational model can be used as the teacher. The student encoder is much smaller than the teacher's ($\sim$5M vs.~$\sim$79M parameters). In the classification heads, the encoder's temporal outputs are combined into invocation-specific representations using attention-summarization (Sum in \cref{fig:main}), followed by linear classifiers and SoftMax. We propose combining knowledge distillation techniques as follows: first, we apply the embedding distillation by computing the MSE loss between the teacher and student encoder outputs ($L_{ED}$). This transfers the teacher's representation knowledge to the student. Second, we introduce the attention regularization to enforce temporal consistency between the two models ($L_{AR}$). Third, in addition to the label-supervised losses ($L_{DDSD}$), we also apply pseudo-labelling ($L_{PL}$), where the teacher model's outputs serve as targets for the student model, aligning their predictions. By combining these losses within aKD, we effectively distill the teacher's expertise into the student. In our experiments, we evaluate the impact of adding the different KD losses, and show on the real-world data the benefits of the aKD approach with those.

The following sections outline the system and our distillation framework design (\cref{sec:overview}). In \cref{sec:kd}, we define each distillation loss, discuss its role, and introduce the aKD. Finally, in \cref{sec:exp}, we show the experimental results and include ablations to explore the effectiveness of different distillation setups.
 
\begin{table*}[t]
\caption{Performance comparison  of various KD approaches, reporting EER in percentages (\%). Large accuracy improvements are observed due to KD, outperforming the student model without KD. The best and second-best performances are denoted by bold and underlined characters, respectively.} 
\centering
\vspace{-.5em}
\addtolength{\tabcolsep}{4pt}
\begin{tabular}{l l |  l l l | l l l}
\toprule
& & \multicolumn{6}{c}{Student Architectures} \\
\cline{3-8} 
& & \multicolumn{3}{c|}{Transformer} & \multicolumn{3}{c}{Conformer}  \\ 
\multicolumn{2}{c|}{Models} & \multicolumn{1}{c}{HAG} & \multicolumn{1}{c}{AG} & \multicolumn{1}{c|}{FCO} & \multicolumn{1}{c}{HAG} & \multicolumn{1}{c}{AG} & \multicolumn{1}{c}{FCO} \\ \hline \hline
\multirow{2}{*}{Baselines} & Teacher & 1.66 & 3.87 & 7.74 & 1.66 & 3.87 & 7.74 \\ 
& DDSD w/o KD & 2.62 & 5.76 & 11.78 & 1.57 & 5.24 & 10.87  \\ \hline
\multirow{4}{*}{Conventional KD} & $\mathcal{L}_{\text{DDSD}} + \mathcal{L}_{\text{ED}}$ & 3.14 & 6.81 & 12.32 & 1.57& 5.24 &9.06 \\
&$\mathcal{L}_{\text{DDSD}} + \mathcal{L}_{\text{ED}} + \mathcal{L}_{\text{AR}}$ &2.09&5.76 &12.32& \textbf{1.05} & 5.24&  9.60 \\
&$\mathcal{L}_{\text{PL}} + \mathcal{L}_{\text{ED}} + \mathcal{L}_{\text{AR}}$ &2.62&5.24&10.33& 1.57& \textbf{4.19}& \textbf{8.51} \\
&$\mathcal{L}_{\text{DDSD}} + \mathcal{L}_{\text{PL}} + \mathcal{L}_{\text{ED}} +\mathcal{L}_{\text{AR}}$ &2.62 &5.24& \underline{10.14}& 1.57&4.71&8.88 \\ \hline
\multirow{4}{*}{Adaptive KD} & $\mathcal{L}_{\text{DDSD}} + \mathcal{L}_{\text{ED}}$ & 3.14 & 6.81 & 13.22 & 1.57 & 6.81 &9.96 \\
&$\mathcal{L}_{\text{DDSD}} + \mathcal{L}_{\text{ED}} + \mathcal{L}_{\text{AR}}$ & \textbf{1.57} & 5.24 & 12.86 & 1.57 & 5.76 & 9.42 \\
&$\mathcal{L}_{\text{PL}} + \mathcal{L}_{\text{ED}} + \mathcal{L}_{\text{AR}}$ & 2.62 &  \textbf{4.71} & \underline{10.14} & 2.09 & 5.23 & 9.24\\
&$\mathcal{L}_{\text{DDSD}} + \mathcal{L}_{\text{PL}} + \mathcal{L}_{\text{ED}} +\mathcal{L}_{\text{AR}}$ & \underline{2.09}&  \underline{4.87}&  \textbf{9.96} & \textbf{1.05} & \textbf{4.19} & \underline{8.70}\\
\bottomrule
\end{tabular}
\label{table:results}
\vspace{-1.5em}
\end{table*}

\vspace{-1em} \section{System Overview}  \label{sec:overview}
\subsection{Device-Directed Speech Detection (DDSD) Model} 
Our DDSD model takes as an input temporal sequence of audio features $X=[x_1,..,x_T]$ and classifies them as device-directed or not $y \in \{0, 1\}$ (see~Fig.~\ref{fig:main}). We design DDSD following the unified acoustic detector architecture proposed in~\cite{rudovic2023uad}. The acoustic features (mel-filter banks from input audio) $X$ are first transformed by the encoder into an embedding sequence $E=[e_1,..,e_T]$ through the acoustic encoder. Then, the audio embedding $E$ is summarized into a single vector $Z$ by a global attention mechanism (GA)~\cite{wang2021global}. Formally, we obtain the GA weights as follows:
\begin{align}
\alpha_t=\frac{exp(s_t)}{\sum_{i\in\{1,..,T\}} exp(s_i)},\,\text{and}\, s_t=e_t\theta,
\label{eq:attn}
\end{align} 
where $t$ is the frame index, and the GA linear projection $\theta$ computes the frame-wise attention weights $\alpha_t$. These are used to obtain the logit $Z=\sum_{i\in\{1,..,T\}} \alpha_i e_i$. Finally, $Z$ is passed through task-specific fully connected layers and SoftMax to produce the final mitigation score $p$, which is the probability of the input audio being device-directed. Finally, the joint loss is defined as cross-entropy (CE) loss:
\begin{align}
\mathcal{L}_{\text{DDSD}} = \text{CE}(y,p) \label{eq:aftm},
\end{align}
where $y$ is a one-hot encoded vector of the ground truth label. As in~\cite{rudovic2023uad}, we adopt multi-task learning that involves training separate adapters for different invocation types alongside shared multi-task learning objectives. In Figure~\ref{fig:main}, the adapters are trained for each invocation type. The invocation types include two keyword-based methods: `Hey Agent' (HAG) and `Agent' (AG), and one keyword-free follow-up conversation (FCO). The teacher and student models share the same architecture for these classification heads.

\vspace{.5em}\noindent \textbf{Acoustic Encoder.}
In this work, we use two types of acoustic encoder for our DDSD model; Transformer \cite{vaswani2017attention} and Conformer\cite{gulati2020conformer} to investigate generalization of the proposed KD architectures. Recently, conformer-based acoustic encoders have demonstrated great performance in speech recognition tasks~\cite{gulati2020conformer}. However, the conformer architecture has yet to be tested for DDSD. Therefore, we aim to explore the advantages of conformers for DDSD. Convolution and self-attention (SA) modules are placed in a serial manner in the Conformer encoder. The convolution module captures local features whereas the SA module captures the global features. We place these modules in parallel instead to capture features at different resolutions \cite{szegedy2014goingdeeperconvolutions, peng2022branchformerparallelmlpattentionarchitectures}.
The SA and Convolution module output are concatenated, and added a bottleneck layer to maintain the hidden size. We provide more details in Sec.~\ref{sec:imple}.

\subsection{Teacher Model}
For our teacher model, we selected a conformer-based model trained for an automatic speech recognition (ASR) task~\cite{yao2021wenet}. We tested several acoustic foundation models to use as a teacher model, but the ASR model outperformed the other candidates in our evaluations (since it was trained on in-domain data), making it the most suitable choice. This teacher model, featuring 12 conformer layers and 79 million parameters, had its encoder frozen. We then appended classification heads identical to those in the DDSD model. These heads were trained on the DDSD learning objective defined in~\cref{eq:aftm}.

\section{Adaptive Knowledge Distillation} \label{sec:kd}

\subsection{Knowledge Distillation} \label{sec:distil_terms}
In our approach, we integrate several distillation losses to transfer knowledge effectively from the teacher to student models. Embedding Distillation (ED) aligns the acoustic representations, encouraging the student models to develop more general representations. This enhances their performance on the DDSD task. Pseudo Labeling (PL) further aids this process by guiding the student model to mimic the DDSD responses of the teacher model. This not only helps the student adapter emulate the teacher adapter but also directly boosts the student model's DDSD capabilities. Moreover, we propose a novel Attention Regularizer (AR), which aligns the temporal similarity between student and teacher representations, facilitating context transfer from the teacher to the student. Throughout the text, the superscripts $\text{S}$ and $\text{T}$ are employed to distinguish between the student and teacher models, respectively.

\vspace{.5em}\noindent\textbf{Embedding Distillation (ED).} We distill the output of the ASR encoder of the teacher model, $Z^\text{T}$, into the acoustic encoder of the student model, $Z^\text{S}$, enabling the student's acoustic encoder to mimic the robust acoustic representation of the teacher's ASR encoder. Since the ASR encoder of the teacher is frozen and not fine-tuned for the DDSD task, it offers a more general representation, beneficial for the DDSD task.
For embedding distillation, we use mean square error:
\begin{align}
\mathcal{L}_{\text{ED}} = \text{MSE}(E^\text{T}, E^\text{S}).
\end{align}

\vspace{.5em}\noindent\textbf{Pseudo Labeling (PL).} To facilitate the student adapter in mimicking the responses of the teacher adapter, we introduce pseudo labeling motivated by \cite{gandhi2023distil}
we replace the ground truth target in the DDSD objective (\cref{eq:aftm}) with the output of the teacher model. After applying argmax to obtain the predicted labels, we convert these to one-hot encoded targets. This pseudo-labeling approach is defined as:
\begin{align}
\mathcal{L}_{\text{PL}} = \text{CE}(\hat{y}, p^\text{S}),
\end{align}
where the pseudo-label $\hat{y}$ is derived by applying argmax to the teacher model's output probability $p^\text{T}$.

\vspace{.5em}\noindent\textbf{Attention Regularization (AR).} We leverage Attention Regularization (AR) to transfer temporal context, which is a critical aspect overlooked by existing methods like ED and PL. Temporal context is pivotal in keyword detection as not all frames contain the target keyword. Some timeframes are more salient than others. To address this limitation, we propose an innovative approach by introducing an additional distillation term. This term regularizes the global attention weights between the student and teacher models, effectively emphasizing the importance of different timeframes during the learning process. 
\begin{align}
\mathcal{L}_{\text{AR}} = \sum_{t\in\{1, \ldots, T\}}(\alpha_t^\text{T} - \alpha_t^\text{S})^2,
\end{align}
where $\alpha_t$ is a GA weight at $t$ defined in \cref{eq:attn}.

\vspace{.5em}\noindent\textbf{Objective.} The final KD objective is a weighted sum of the above terms and a DDSD objective:
\begin{align}
    \mathcal{L}_{\text{student}} = \mathcal{L}_{\text{DDSD}} + \lambda_{\text{ED}} \mathcal{L}_{\text{ED}} + \lambda_{\text{PL}} \mathcal{L}_{\text{PL}} + \lambda_{\text{AR}} \mathcal{L}_{\text{AR}},
  \label{eq:objective}
\end{align}
where $\lambda_{\text{ED}}$, $\lambda_{\text{PL}}$, and $\lambda_{\text{AR}}$ are scalar weights for the ED, PL, and AR losses respectively. We provide details on how we tune these weights in \cref{sec:imple}.

\subsection{Adaptive Knowledge Distillation}
We propose Adaptive Knowledge Distillation (aKD), a novel method that trains the teacher adapter and the student model simultaneously. Knowledge distillation between domains can be challenging, as it often requires fine-tuning the entire teacher model for the target task to bridge the domain gap. Conventional KD addresses this by fine-tuning a domain-specific adapter on the downstream domain without the computational burden of fine-tuning the full teacher model. However, it involves two steps:
\begin{enumerate}
\item The teacher model is fine-tuned on the downstream domain with an adapter, while the encoder is frozen.
\item Both the teacher encoder and adapter are then frozen, and knowledge is distilled to the student model.
\end{enumerate}
aKD differs from conventional KD in that it trains the teacher adapter and the student model simultaneously. This process aligns the student and teacher adapters since they are trained together for the same target. Consequently, their knowledge gap is minimal at the beginning of training and gradually reduces throughout the process through KD. In contrast, conventional knowledge distillation leads to a substantial knowledge gap that is difficult to reduce because the teacher was already trained, while the student was not, when the distillation process began. By eliminating the need for the two-step process, aKD streamlines the training process.
\begin{figure*}[t]
    \centering
    \includegraphics[width=.33\linewidth]{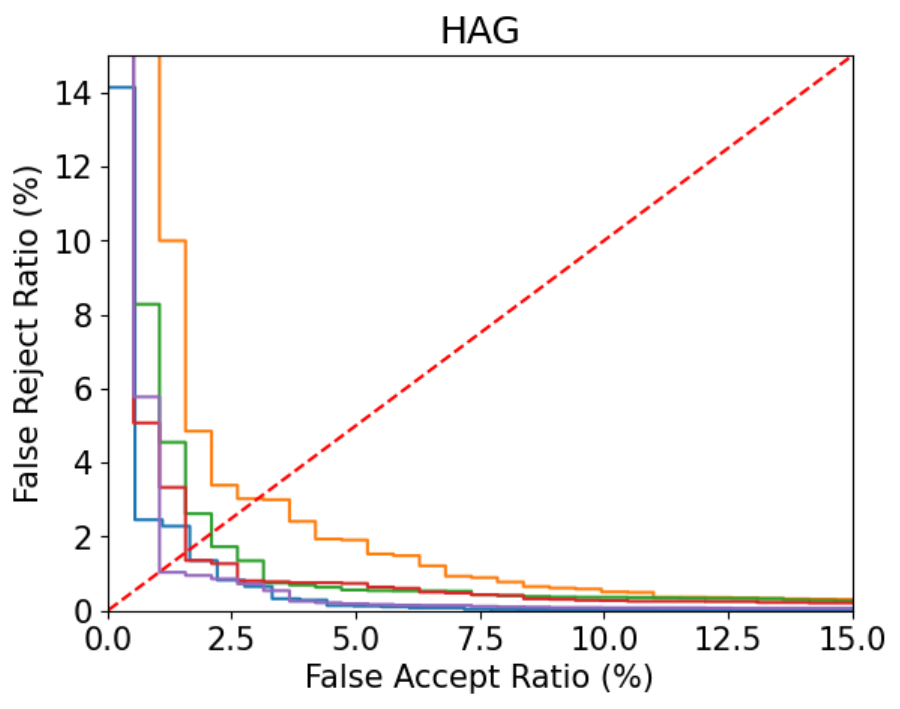}
    \includegraphics[width=.33\linewidth]{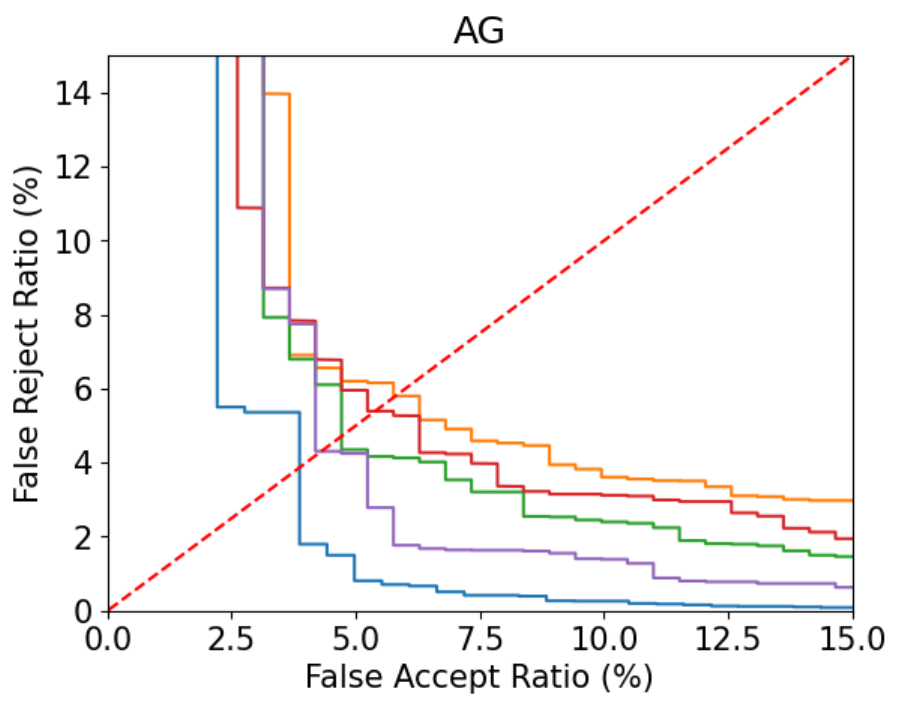}
    \includegraphics[width=.33\linewidth]{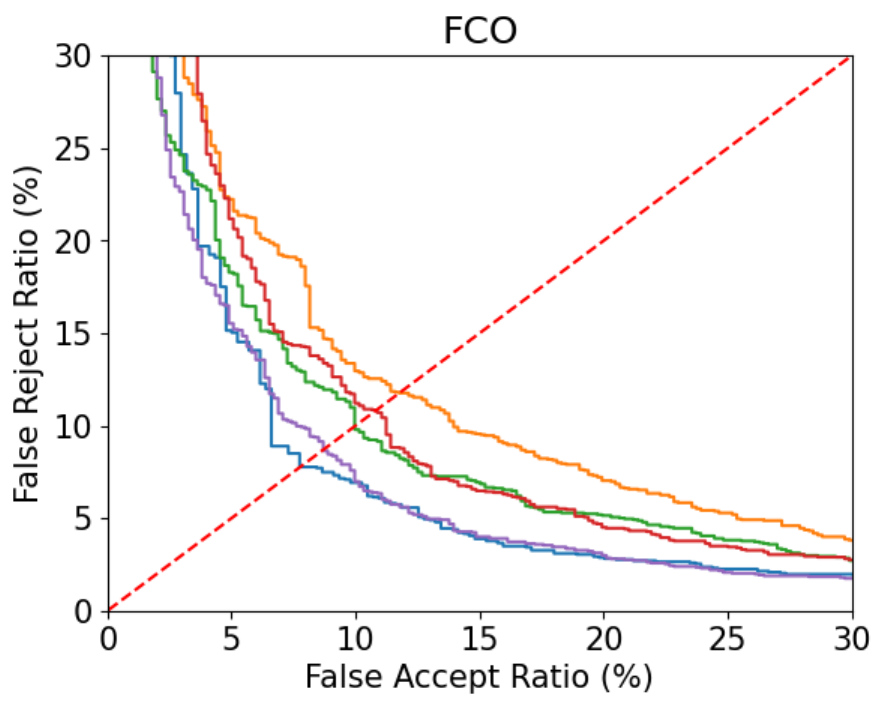}
    \includegraphics[width=1.\linewidth]{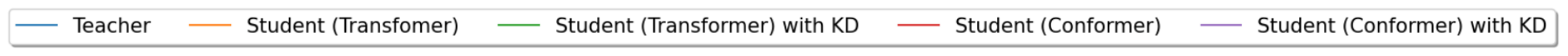}
    \vspace{-1.5em}
    \caption{DET curves on different invocation types. The red dotted line indicates the EER line, where the False Reject Ratio (FRR) and False Accept Ratio (FAR) are equal. The proposed aKD method was used for this plot.}
    \label{fig:det}
    \vspace{-1em}
\end{figure*}

\section{Experiments} \label{sec:exp}
\subsection{Implementation Details} \label{sec:imple}
Each model is trained on 32 $\times$ NVIDIA V100 40GB GPUs for 100 epochs with an batch size of 256.
We employ the Adam optimizer \cite{kingma2014adam} with an initial learning rate of $1 \times 10^{-7}$. To further enhance convergence, we implement an adaptive learning rate scheduler that reduces the learning rate when a validation metric plateaus.
For the student loss in Eq. \ref{eq:objective}, we set the weights as $\lambda_{\text{ED}} \!=\! 1 \times 10^{2}$, $\lambda_{\text{PL}} \!= \!1 $, and $\lambda_{\text{AR}} \!= \!1$. We conducted a grid search of the weights on our held-out validation set. As input, 40-D Mel-filterbank features at 100 fps are used, and the current frame is augmented with 6 neighbouring frames, resulting in 280-D features. For the transformer-based student model, the acoustic encoder is comprised of 8 transformer encoder blocks with 256 hidden units, 4 heads within the SA module, and 1,024 hidden units in the feed-forward layer. For conformer-based student model, we use 8 layers of conformer encoder blocks, where each block has the SA module with 168 hidden units and 4 heads, a convolution module with kernel size 31, and Macaron-style feed-forward layers with 672 hidden units. The model sizes for both Transformer and Conformer based student models are similar, both being around 5M parameters (compared to 79M parameters in the teacher).

\subsection{Dataset and Evaluation Metrics}
We used in-house collected training data, that consists of 2.7K hours of audio samples which includes the three invocation types and users interacting with VA in different contexts, as in~\cite{rudovic2023uad,wagner2025selma}. These were further augmented with room impulse responses and echo residuals, as in previous works \cite{wagner2025selma, rudovic2023uad, higuchi2024multichannel}. To evaluate our model, we also utilized in-house evaluation sets tailored to each invocation type. We report Equal-error-rate (EER) per invocation type. We established two baseline teacher models trained from scratch following Step 1 of conventional KD described in \cref{sec:kd}, serving as the upper bound, and DDSD model trained from scratch, representing our lower bound. Since the EER only provides a single representative value, we further plot the Detection Error Tradeoff (DET) curves in Fig.~\ref{fig:det} to investigate the impact of distillation on different False Accept Rate (FAR) thresholds.


\subsection{Results} \label{sec:results}

We report all our experimental results on DDSD in Table~\ref{table:results}. The table presents results for various student architectures, invocation types, and KD methods. We first see that KD based results bring the performance boost across both Transformer and Conformer architectures compared to the DDSD model without KD. Notably, the accuracy improvement of aKD with the Conformer architecture is substantial for keyword-based invocations (HAG and AG), where aKD achieves an average relative gain of 26\%. Additionally, the relative improvement in keyword-free invocations (FCO) is 19\%.
It highlights the advantages of knowledge distillation for DDSD task. In the following section, we delve into the components that contribute to the performance boost.

\subsubsection{aKD vs Conventional KD}
We begin by comparing conventional KD with our proposed aKD. aKD consistently outperforms conventional KD across both transformer and conformer student architectures and all invocation types. For example, with a conformer-based student model and all KD losses utilized, aKD substantially outperforms conventional KD in EER. It shows improvements of 0.52\% (33\% relative) on HAG, 0.52\% (11\% relative) on AG, and 0.18\% (2\% relative) on FCO. One possible explanation for these results is that when both the teacher and student adapters are optimized jointly, knowledge is progressively distilled into the student model during the teacher's training. This approach effectively reduces the gap between the models over time. In contrast, conventional KD methods often lead to a knowledge gap between the student and teacher models at the start of training, which can hinder the learning process.

\subsubsection{KD Losses}
Next, we delve into the contributions of each distillation term outlined in \cref{sec:distil_terms}. We found that the model utilizing all distillation losses achieves the best or the competitive performance. Comparing models trained with and without $\mathcal{L}_{\text{AR}}$, we demonstrate the performance gain, highlighting $\mathcal{L}_{\text{AR}}$'s temporal-wise context distillation ability. Furthermore, replacing the true targets of $\mathcal{L}_{\text{DDSD}}$ in Eq \ref{eq:aftm}  with pseudo-labels in $\mathcal{L}_{\text{PL}}$ leads to performance improvements, particularly for AG and FCO, though it degrades performance on HAG. Interestingly, combining $\mathcal{L}_{\text{DDSD}}$ and $\mathcal{L}_{\text{PL}}$ results in the best performance, underscoring their complementary nature. Since there are inherent relations between these losses, using all losses together brings more robustness to the model training overall. Further optimizations can be made by a more exhaustive search of the loss weights in the final loss. 

\subsubsection{Conformer VS Transformer Student Models}
To demonstrate the generalizability of our distillation framework, we compare the distillation results of two different student architectures. We first investigate the effectiveness of KD on a Conformer model for the DDSD task, a model architecture that has not been previously reported for this task. When training models from scratch, Conformer shows superior performance compared to the Transformer-based model. To further investigate impact of KD on conformer, we apply the same distillation technique used for the Transformer to the Conformer. As expected, KD improves the performance of both models. The aKD approach consistently outperforms conventional KD, enhancing the performance of both architectures. Interestingly, we observe that the Conformer architecture outperforms the Transformer-based one, aligning with the findings of previous work \cite{gulati2020conformer} on the advantages of Conformers in the acoustic domain. Additionally, we compare the DET curves between Transformer and Conformer-based student models in \Cref{fig:det}. The results show that applying aKD improves the DET curves, reducing the area under the curves for both models across all invocation types.

\section{Conclusion} \label{sec:conclusion}
This paper addresses the challenge of improving device-directed speech detection accuracy in on-device architectures with limited computing resources. Specifically, we introduced an adaptive knowledge distillation approach that transfers effectively the knowledge from the teacher to student. In this approach, we (i) unified the distillation losses focusing on the score, representation and attention distillation, quantifying the impact of each loss on DDSD accuracy, and showing their generalization across different model architectures (attention-only and conformer-based). (ii) We tackled a challenging problem of multi-invocation DDSD and showed that our approach reduces EER by an average of 22\% across different invocation types, and by 22.5\% overall across transformer and conformer-based student architectures and invocations. This makes it a valuable solution for enhancing the user experience of voice-activated devices while being able to deploy it on device due to the small model size. While in this work we did not explore using other teacher models, that is part of our ongoing research.

\bibliographystyle{IEEEtran}
\bibliography{mybib}

\end{document}